\documentclass[sigconf]{acmart}
\AtBeginDocument{%
  \providecommand\BibTeX{{%
    \normalfont B\kern-0.5em{\scshape i\kern-0.25em b}\kern-0.8em\TeX}}}

\acmConference[arXiv]{}{June}{2024}
\begin{document}

\title{Domain-Driven Design Representation of Monolith Candidate Decompositions Based on Entity Accesses}


\author{Miguel Levezinho}
\email{miguel.levezinho@tecnico.ulisboa.pt}
\affiliation{%
  \institution{University of Lisbon Instituto Superior Técnico}
  \city{Lisbon}
  \country{Portugal}
}

\author{Stefan Kapferer}
\email{stefan.kapferer@ost.ch}
\affiliation{%
  \institution{OST Eastern Switzerland University of Applied Sciences}
  \city{Rapperswil}
  \country{Switzerland}
}

\author{Olaf Zimmermann}
\email{olaf.zimmermann@ost.ch}
\affiliation{%
  \institution{OST Eastern Switzerland University of Applied Sciences}
  \city{Rapperswil}
  \country{Switzerland}
}

\author{António Rito Silva}
\email{rito.silva@tecnico.ulisboa.pt}
\affiliation{%
  \institution{University of Lisbon Instituto Superior Técnico}
  \city{Lisbon}
  \country{Portugal}
}


\renewcommand{\shortauthors}{Levezinho et al.}

\begin{abstract}
Microservice architectures have gained popularity as one of the preferred architectural approaches to develop large-scale systems, replacing the monolith architecture approach. Similarly, strategic Domain-Driven Design (DDD) gained traction as the preferred architectural design approach for the development of microservices. However, DDD and its strategic patterns are open-ended by design, leading to a gap between the concepts of DDD and the design of microservices. This gap is especially evident in migration tools that identify microservices from monoliths, where candidate decompositions into microservices provide little in terms of DDD refactoring and visualization. This paper proposes a solution to this problem by extending the operational pipeline of a multi-strategy microservice identification tool, called Mono2Micro, with a DDD modeling tool that provides a language, called Context Mapper DSL (CML), for formalizing the most relevant DDD concepts. The extension maps the content of the candidate decompositions, which include clusters, entities, and functionalities, to CML constructs that represent DDD concepts such as \textit{Bounded Context}, \textit{Aggregate}, \textit{Entity}, and \textit{Service}, among others. The results are validated with a case study by comparing the candidate decompositions resulting from a real-world monolith application with and without CML translation.
\end{abstract}


\keywords{Domain-Driven Design, Microservices, Migration.}



\maketitle

\section{Introduction}

Microservice architectures have become one of the architectures of choice for emerging large enterprise applications~\cite{Hanlon06,Francesco18}. This adoption results from the advantages of partitioning a large system into several independent services, which provide qualities such as strong boundaries between services; independent development, testing, deployment, and scaling of each service; and service-tailored infrastructures~\cite{FowlerMicroservices15,FowlerTradeOffs15,Richardson17}. On the other hand, topics such as how to distribute the system and the consistency model might stagger the design early on. The use of a monolith architecture, where the business logic of the system is interconnected, has the advantage that it does not require early modularization. The neat identification of modules occurs through refactorings, after initial development, which allows one to explore the application domain first~\cite{FowlerMonoFirst15}.

Therefore, it is common practice to start with a monolith and, as the system grows in size and complexity, migrate to a more modular architectural approach, such as a modular monolith~\cite{Hayhood17} or a microservice architecture. Since this architectural migration is not trivial~\cite{Ponce19}, recent research has proposed approaches and tools to help the migration process~\cite{Abdellatif21,Abgaz23}.

This has led to the development of Mono2Micro, a modular and extensible tool for the identification of microservices in a monolith system~\cite{Lopes23}. Mono2Micro focuses on identifying transactional contexts to inform its generated candidate decompositions~\cite{Nunes19}. To this end, it integrates several approaches, such as static code analysis of monolith accesses to domain entities~\cite{Santos22}, dynamic analysis of monolith execution logs~\cite{Andrade23}, lexical analysis of abstract syntactic trees of monolith methods~\cite{Faria23}, and analysis of the history of monolith development~\cite{Lourenco23}. Furthermore, Mono2Micro supports a set of measures and graph views to evaluate the quality of the generated candidate decompositions~\cite{Santos20}.

However, as with most research on the identification of microservices in monolith systems, Mono2Micro does not allow software architects to further model generated candidate decompositions using Domain-Driven Design (DDD)~\cite{Evans03}, which has shown good results on microservice design~\cite{Vural21} and growing interest in the industry~\cite{Ozkan23}. Instead, Mono2Micro representations of candidate decompositions are based on sequences of read and write accesses to the monolith domain entities, which are difficult to work with when trying to redesign the original monolith system and its functionalities for a modular architecture.

This paper addresses this problem by providing a representation of the Mono2Micro candidate decompositions in terms of automatically generated tactical and strategic DDD patterns. In this way, software architects can work on candidate decompositions from the perspective of DDD.

This is achieved by extending the operational pipeline of \\Mono2Micro with a connection to Context Mapper, a DDD-focused modeling tool that provides a Domain-Specific Language, named Context Mapper DSL (CML). CML supports the declarative description of DDD domain models, using DDD concepts as building blocks of the language~\cite{Kapferer20_dsl}. With this goal in mind, the following research questions are raised:


\begin{itemize}
    \item \textbf{RQ1}: How can current approaches to the identification of microservices in monolith systems be extended to include DDD?
    \item \textbf{RQ2}: Can the results of a candidate decomposition based on entity accesses be represented in terms of DDD?
    \item \textbf{RQ3}: Can an architect benefit from the use of a tool that integrates DDD when analyzing and working on a candidate decomposition?
\end{itemize}

To answer these research questions, a real monolith system was used as a case study. The resulting candidate decompositions of this system were generated with and without the new DDD modeling capabilities and then compared.

The remainder of this paper is structured as follows. Section~\ref{sec:related_work} goes over the current literature on DDD application and microservice identification tools. Section~\ref{sec:background} gives some background on the Mono2Micro and Context Mapper tools. Section~\ref{sec:solution_architecture} presents the solution to the aforementioned research questions. Section~\ref{sec:case_study} provides the validation of the solution with a case study application, and in Section~\ref{sec:discussion} the results and answers to the research questions are discussed. Finally, Section~\ref{sec:conclusion} concludes the paper.

\section{Related Work}
\label{sec:related_work}

The application of DDD in microservice development, although widely practiced, is still poorly formulated~\cite{Singjai21}. Research on DDD modeling techniques is still sparse, especially in terms of modeling tools that leverage tactic and strategic DDD patterns~\cite{Ozkan23}.

Most research takes advantage of already developed models and diagram standards in the industry to convey DDD concepts. The use of annotated constructs is one of the most common approaches, such as in~\cite{Rademacher17}, where a mapping from DDD to UML is presented with the use of annotations inside UML class constructs, or in~\cite{MinhLe18}, where an annotation-based DSL was developed to scope objects and attributes within the concepts of DDD. However, they do not support all DDD patterns, especially strategic ones such as \textit{Bounded Context} relationships, which are useful when modeling microservices from candidate decompositions.

Context Mapper is an exception to this, providing a DSL to model tactic and strategic DDD patterns~\cite{Kapferer20_dsl}, rapid model prototyping by deriving \textit{Domains} and \textit{Bounded Contexts} from use case definitions~\cite{Kapferer21_dda}, and integration with other technologies such as Microservice Domain-Specific Language (MDSL)~\cite{Kapferer20_dds}.

Other research also explores the extensibility of DDD to better fit other stages of software development. In~\cite{Hippchen17} they define \textit{Domain Views}, which enable different stakeholders to perceive the domain model with their respective knowledge base. The Context Mapper tool also provides \textit{Domain Views} through the definition of types of \textit{Bounded Context} and \textit{Context Maps}~\cite{Kapferer20_dsl}.

On the other hand, there has been extensive and recent research on the identification of microservices in monolith systems~\cite{Abdellatif21}. These approaches provide a rich set of decomposition criteria and metrics to assess the generated decompositions. In particular, Mono2Micro~\cite{Lopes23} is a modular and extensible tool for those criteria and methods. However, these tools do not provide output that enables DDD-based editing and modeling, they mostly provide decompositions that are service-oriented and not domain-oriented.

To our knowledge, the only tool that supports the reverse engineering of DDD concepts is the Discovery Library tool~\cite{Kapferer20_dsl}, which provides a way to reverse engineer domain models from Spring Boot\footnote{\url{https://spring.io/projects/spring-boot}} service APIs using discovery strategies. From the analysis of the code, it finds specific Spring Boot annotations and maps them to the corresponding DDD concepts. It generates \textit{Bounded Contexts} from @SpringBootApplication annotated classes and \textit{Aggregates} from @RequestMapping annotated classes.

\section{Background}
\label{sec:background}

To better inform the integration of Context Mapper into the \\Mono2Micro pipeline, this section gives an overview of the architecture of both tools and compares them.

\subsection{Mono2Micro}

Mono2Micro is a migration tool that provides candidate monolith decompositions composed of clusters of domain classes. This work initially focused on the identification of microservices driven by the identification of transactional contexts~\cite{Nunes19}, but other strategies have been added~\cite{Andrade23,Faria23,Lourenco23}.

\begin{figure*}[htb]
    \centering
    \includegraphics[width=0.8\textwidth]{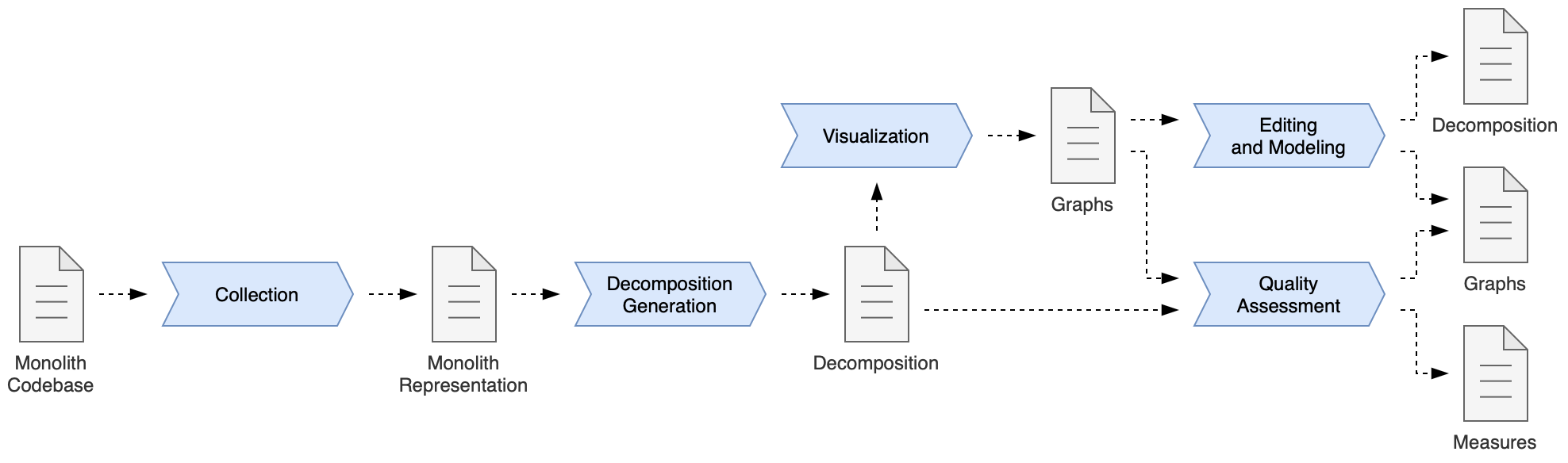}
    \caption{The five stages of the Mono2Micro operational pipeline~\cite{Lopes23}. Each stage can use the output of former stages as input.}
    \label{fig:m2m_pipeline}
\end{figure*}

Mono2Micro is designed as a pipeline, which is represented in Figure~\ref{fig:m2m_pipeline}. The five stages of the pipeline are:

\begin{enumerate}
    \item \textit{Collection}: Implements several static and dynamic code collection strategies to represent monoliths, including representations based on accesses to source code domain entities, functionality logs, and commit history and authors.
    \item \textit{Decomposition Generation}: Partitions the monolith domain entities into clusters using a set of similarity criteria, with a focus on producing good quality decompositions.
    \item \textit{Quality Assessment}: Compares the decompositions and calculates the measures that are used to evaluate the generated decompositions. The measures include coupling, cohesion, size, and complexity.
    \item \textit{Visualization}: Depicts decompositions in the form of graphs with multiple levels of detail. Nodes and edges can represent different elements, depending on the chosen collection strategy.
    \item \textit{Editing and Modeling}: Provides an interface with operations to modify the automatically generated decompositions so that the architect can refine them. Quality measures are also automatically recalculated, if applicable.
\end{enumerate}

Each stage is composed of one or more modules that output artifacts for the next stage in the pipeline. The underlying model of the tool that makes up these modules and artifacts is also built with several extension points, making it possible to support multiple decomposition strategies.

However, this pipeline does not include any way for an architect to model candidate decompositions using DDD after the \textit{Decomposition Generation} stage. More concretely, in the \textit{Visualization} stage, graph representations of the decomposition include cluster-based views of the decomposition domain entities, and functionality-based views that represent its sequence of accesses to domain entities ("Graphs" in Figure~\ref{fig:m2m_pipeline}). There are no DDD-based views that show the model of each candidate microservice. Likewise, the \textit{Editing and Modeling} stage does not contain any operations related to the application of DDD. This is where Context Mapper comes in.

\subsection{Context Mapper}

\begin{figure}[htb]
    \centering
    \includegraphics[width=8.5cm]{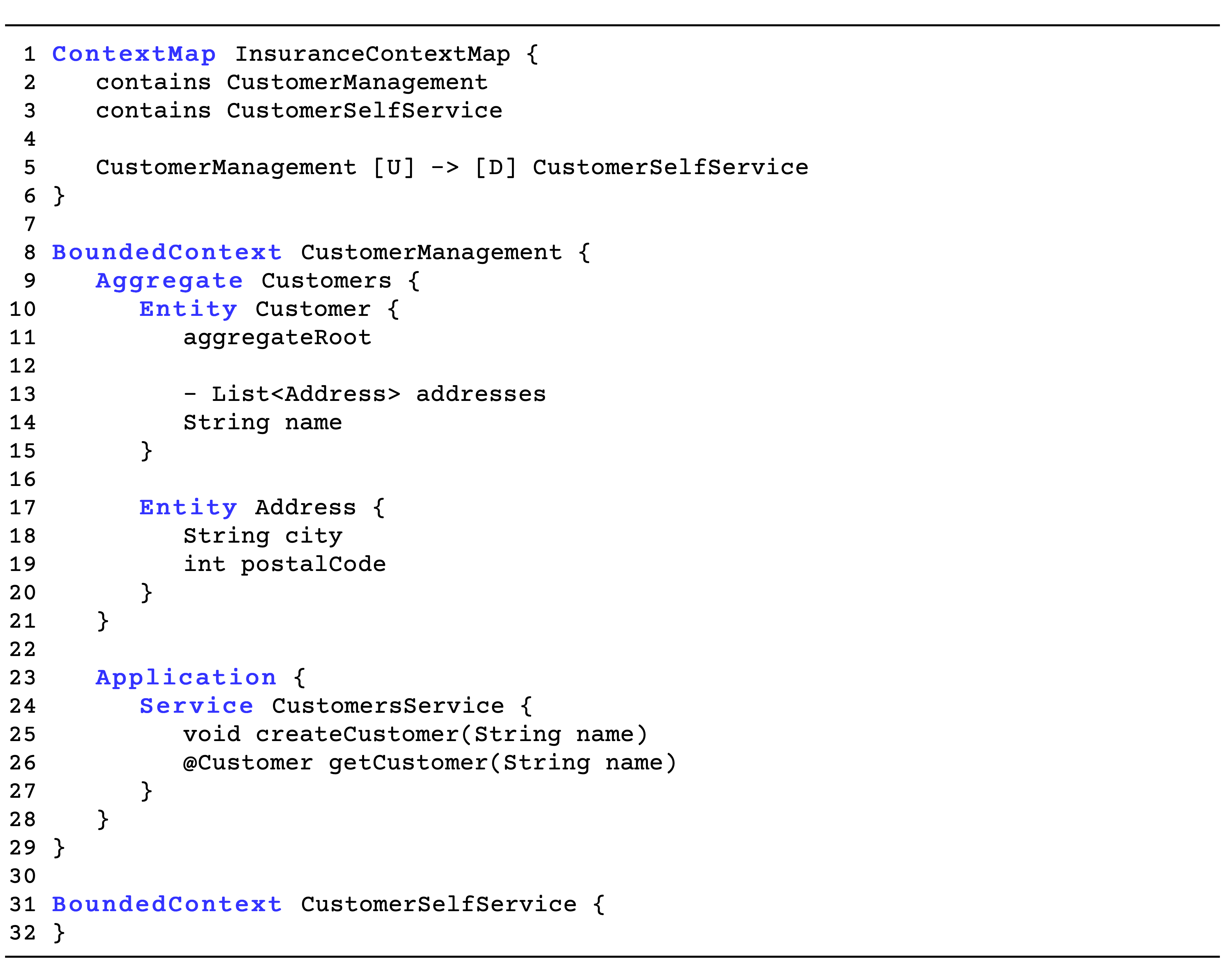}
    \caption{Example syntax of CML, containing the syntax for defining a \textit{Context Map} (1-6); \textit{Bounded Contexts} (8-29, 31-32); \textit{Aggregates} (9-21); \textit{Entities} (10-15,17-20); and \textit{Services} (24-27).}
    \label{fig:cml_example}
\end{figure}

Context Mapper is a modeling framework that provides a DSL to design systems using DDD concepts. This DSL, henceforth called Context Mapper DSL (CML), was developed to unify the many patterns of DDD and their invariants in a concise language~\cite{Kapferer20_dsl}. Figure~\ref{fig:cml_example} shows an example of the CML syntax, with the declaration of a \textit{Context Map} containing two \textit{Bounded Contexts}.

Within \textit{Bounded Contexts}, one can define \textit{Aggregates}, which consist of a group of closely related domain objects that form a unit for the purpose of data consistency. This consistency is enforced inside the \textit{Aggregate} by its root \textit{Entity}, which represents the only entry point. For example, in Figure~\ref{fig:cml_example} the \texttt{Customers} aggregate has the \texttt{Customer} entity as its root.

Although DDD focuses on the \textit{Domain Layer} of systems, where the business logic is residing, a CML \texttt{Bounded Context} can also represent the \textit{Application Layer}, which manages services that call different parts of the system, including processes in other layers. Using the \texttt{Application} keyword, \textit{Application Services} can be defined, among other constructs, and contain operations like \\\texttt{createCustomer} and \texttt{getCustomer} as represented in Figure~\ref{fig:cml_example}.

In addition to the CML language, Context Mapper also contains other utilities to facilitate modeling activities. These include the following:

\begin{enumerate}
    \item \textit{Discovery Library}: Implements several strategies to reverse engineer source code artifacts and represent them in CML~\cite{KapfererCM}.
    \item \textit{Architectural Refactoring}: Includes operations to refactor and transform CML code for easier modeling.
    \item \textit{Diagram Generators}: Provide translators to visualize CML artifacts in diagram form, such as UML representations of \textit{Bounded Contexts} and BPMN maps of \textit{Aggregate} states.
\end{enumerate}

Each of these features has similarities with the features in \\Mono2Micro. First, the Discovery Library performs a similar job as the Collectors of Mono2Micro, but more importantly, it provides a way to generate CML from its input. Second, the Architectural Refactoring (AR) module supports the architect on the edition and modeling of CML models, as the Editing and Modeling stage of Mono2Micro. However, AR operations are built on DDD concepts. Finally, the Diagram Generators module can provide ways to view a candidate decomposition from the perspective of DDD, also something missing in Mono2Micro, which presents decompositions as a graph of clustered domain entities.

\section{Solution Architecture}
\label{sec:solution_architecture}

\begin{figure*}[htb]
    \centering
    \includegraphics[width=0.7\textwidth]{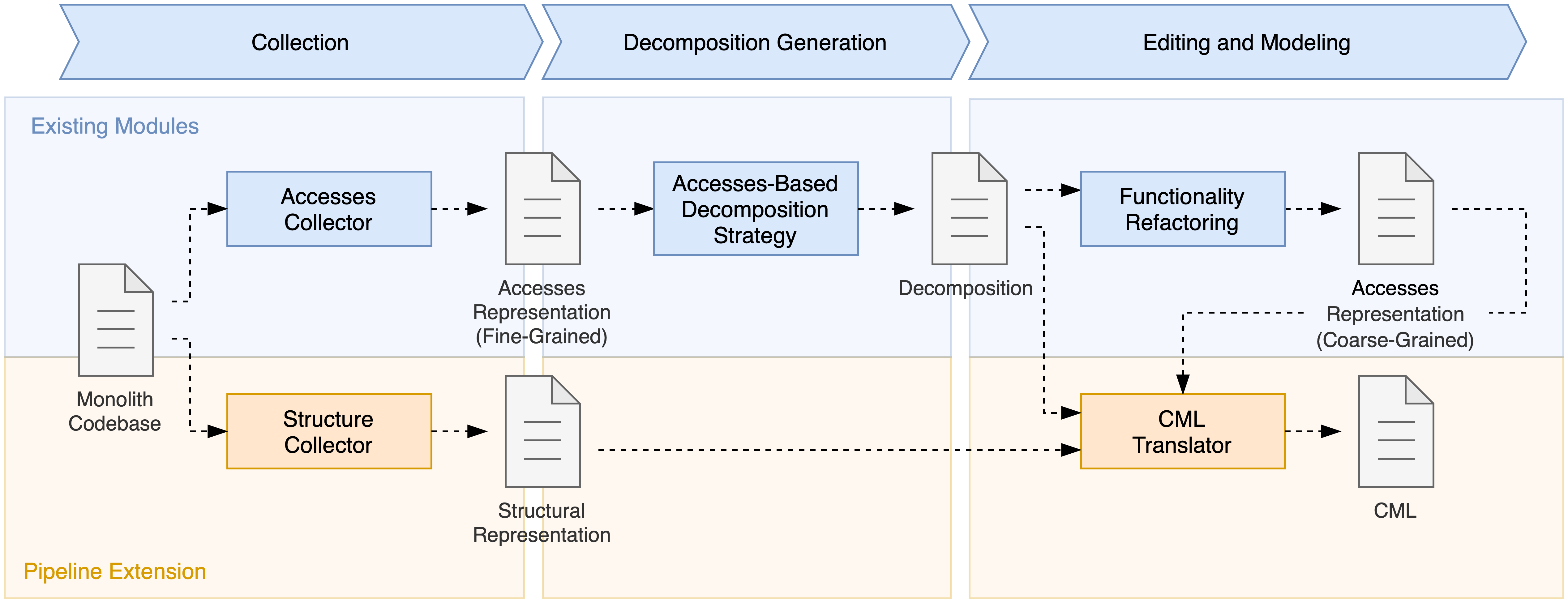}
    \caption{Mono2Micro pipeline extension to support CML representation of candidate decompositions. In blue, the relevant pipeline steps (top) and modules. In orange, the extension to the pipeline, composed of the addition of new modules.}
    \label{fig:m2m_pipeline_extension}
\end{figure*}

The proposed extension to the Mono2Micro microservice identification pipeline provides a representation of candidate decompositions in CML, so that DDD can be used for modeling and refactoring activities. Figure~\ref{fig:m2m_pipeline_extension} shows this extension in terms of modules and their input and output artifacts. The top process bar represents the relevant stages of the Mono2Micro pipeline, and the different colors separate existing modules from new ones. The following sections, each corresponding to one of the research questions, explain each module and artifact in more detail.

\subsection{Tool Integration}
\label{subsec:tool_integration}

Mono2Micro and Context Mapper are built with an emphasis on modularity and extensibility. This makes it viable for Context Mapper to integrate into the Mono2Micro pipeline. However, it is still important to respect the models of each tool to avoid compromising their internal cohesion. In practice, this meant pursuing a low-coupling solution when connecting the tools. This solution was achieved by leveraging on the Discovery Library (DL).

As described in Section~\ref{sec:background}, the DL is a standalone tool capable of generating CML code. This is done using discovery strategies that translate input into CML. Since the DL was designed to be highly extensible, it also provides an API for the creation of these discovery strategies. Using this API, the Mono2Micro pipeline was extended with a module that defines new discovery strategies capable of translating candidate decompositions into CML. This module is represented by the \textit{CML Translator} in Figure~\ref{fig:m2m_pipeline_extension}.

The \textit{CML Translator} has two stages. In the first, the internal representations of a decomposition in the Mono2Micro model are used to create a JSON contract that contains all the information needed to map a candidate decomposition to CML. This contract serves as input for the new discovery strategies and adds a layer of decoupling between the Mono2Micro model and the DL model, ensuring that changes made to the former do not inadvertently propagate to the latter. In the second stage, the new discovery strategies translate the contract to an internal representation of CML in the DL model. This model is, in turn, automatically converted to actual CML code.

\begin{figure}[htb]
    \centering
    \includegraphics[width=5.5cm]{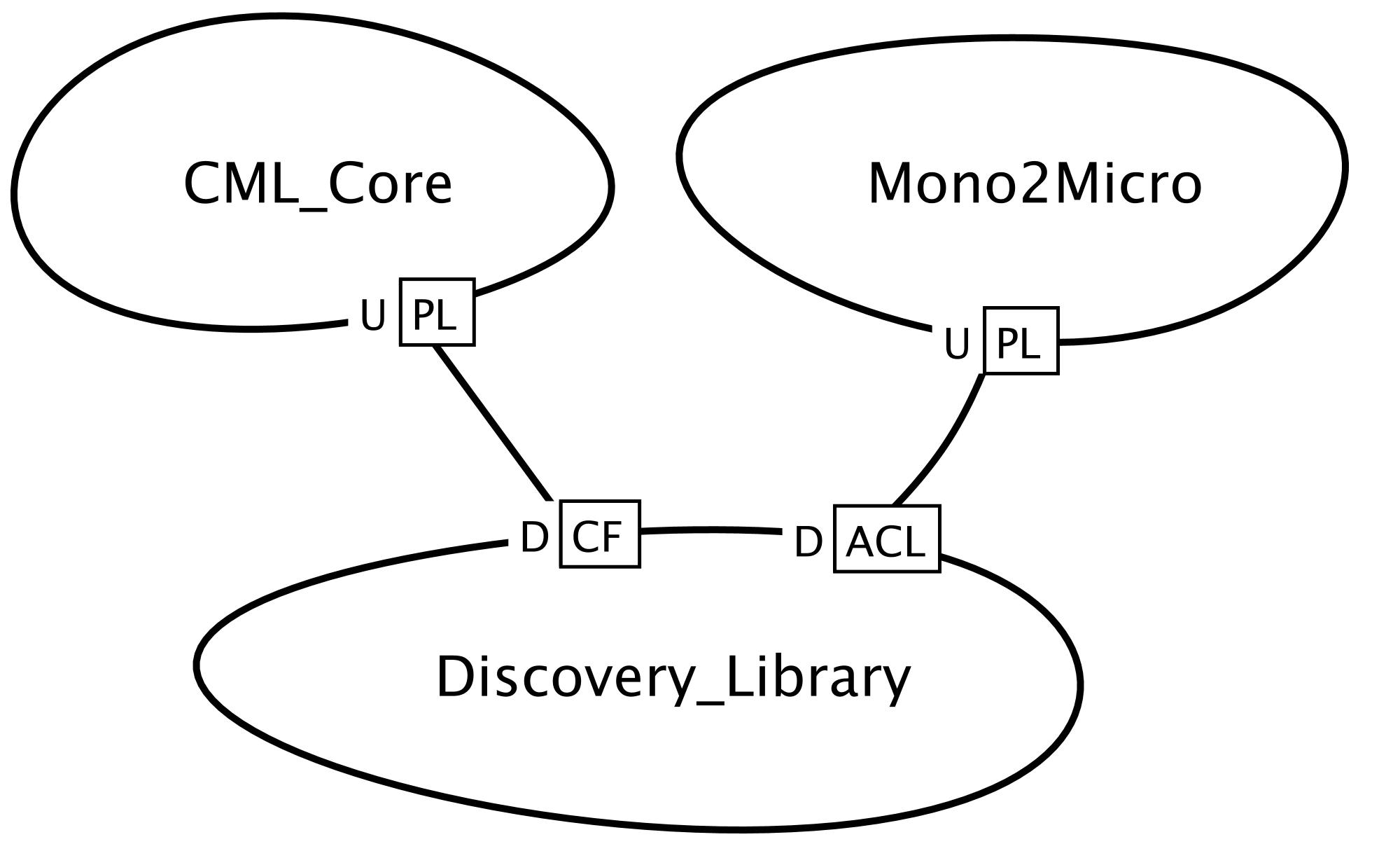}
    \caption{Generated Context Map diagram of the integration between Mono2Micro and Context Mapper. PL stands for \textit{Published Language}, ACL stands for \textit{Anti-Corruption Layer}, and CF stands for the \textit{Conformist} pattern.}
    \label{fig:tool_integration}
\end{figure}

Figure~\ref{fig:tool_integration} shows how DL is used in this process from the perspective of DDD. On the right, the DL acts as an \textit{Anti-Corruption Layer}, protecting its internal model from the Mono2Micro model present in the inbound contract. On the left, the DL model aligns closely with the CML model to facilitate the generation of CML, represented by the use of the \textit{Conformist} pattern.

\subsection{DDD Mapping}
\label{subsec:ddd_mapping}

For the new discovery strategies to perform the translation to CML, the concepts that form a candidate decomposition must be mapped to the DDD concepts first. Since DDD and its concepts are structural in nature~\cite{Evans03}, a candidate decomposition was also structurally defined, based on its internal representation in Mono2Micro.

A candidate decomposition is composed of three key concepts: \textbf{entities}, which represent domain classes in the monolith; \textbf{clusters}, which represent a set of entities grouped by similarity criteria through a clustering algorithm; and \textbf{functionalities}, which represent sequences of read/write accesses to entities in one or more clusters.

\begin{figure}[htb]
    \centering
    \includegraphics[width=8.5cm]{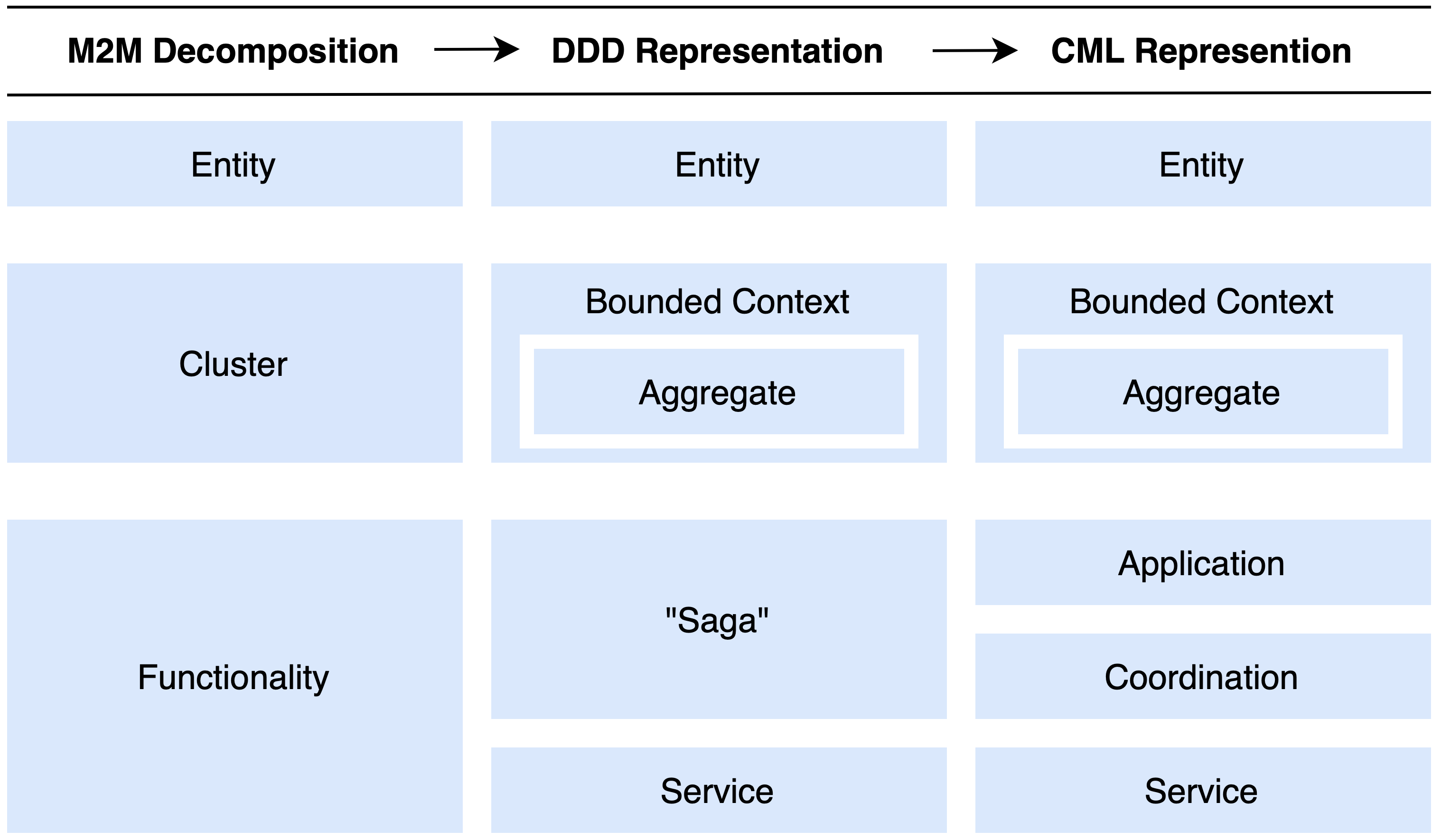}
    \caption{Mapping strategy of candidate decomposition concepts from Mono2Micro (M2M) to DDD and CML.}
    \label{fig:translation_strategy}
\end{figure}

Mapping a candidate decomposition to DDD corresponds to mapping these three concepts and understanding what information is needed from Mono2Micro once a DDD concept is chosen. Figure~\ref{fig:translation_strategy} shows a summary of the achieved mappings, which are discussed in the next paragraphs.

\subsubsection{Entity Mapping} The entities of a candidate decomposition, by definition, are already based on the concept of \textit{Entity} from DDD, which facilitates this mapping. The main difference is that Mono2Micro does not require the internal structure of entities to generate candidate decompositions, while in DDD and CML the internal state and relationships with other entities are relevant information to model an \textit{Entity}.

To guarantee a more complete translation of candidate decomposition entities into CML, a new source code collector module was added to the Mono2Micro Collection stage, aptly named \textit{Structure Collector} as shown in Figure~\ref{fig:m2m_pipeline_extension}. This module uses the Spoon Framework library~\cite{Spoon15} to analyze and collect structural information from entities in the monolith domain, including entity names, entity attributes, and relationships between entities, i.e. composition or inheritance. 

\subsubsection{Cluster Mapping} The main criteria that dictate how entities are clustered in the Mono2Micro Decomposition stage are based on transactional similarity. This means entities commonly accessed together (i.e. read/write) during the same transactions are more likely to belong in the same cluster. Similarly, a DDD \textit{Aggregate} is defined as a group of tightly coupled domain objects that can be seen as a unit for the purpose of data changes during transactions, which makes it a good fit to represent a cluster.

However, the concept of cluster also fits the concept of a DDD \textit{Bounded Context}. This is because clusters define physical boundaries between microservices in a candidate decomposition and can be evaluated based on coupling with other clusters in the same decomposition. This dual mapping of the cluster concept could be achieved with different variations in the number of generated \textit{Bounded Contexts} and \textit{Aggregates}, but in the end the chosen mapping was to take each cluster and generate a corresponding \textit{Bounded Context} and single \textit{Aggregate} inside it, which in turn contains all the entities in the cluster.

Compared to other mapping combinations, this solution has several advantages. To start with, it satisfies all the given definitions of a cluster in Mono2Micro. In contrast, solutions such as mapping each cluster to an \textit{Aggregate} and adding all \textit{Aggregates} to a single \textit{Bounded Context} would remove the distributed nature on which the decomposition is based, and mapping each cluster to a single \textit{Bounded Context} while defining one \textit{Aggregate} per \textit{Entity} would completely ignore the transactional similarity that exists between entities on a cluster. This does not mean that the end product is to have one \textit{Aggregate} per \textit{Bounded Context}. It is important to remember that the generated CML code is by no means final and that further refactoring is expected by the architect doing the modeling. Starting from this initial mapping that satisfies the definition of a cluster, architects have the ability to further refine the model by partitioning the generated \textit{Aggregate} of each \textit{Bounded Context} using not only entity access information, but also the new structural context of entities that is not available in Mono2Micro.

On the subject of this structural context, there is also a caveat to take into account. Entities that share structural relationships in the monolith are likely to end up in different clusters after decomposition. This makes it necessary to ensure that the generated CML code does not contain \textit{Entities} directly referencing other \textit{Entities} in outer \textit{Bounded Contexts}.

\begin{figure}[htb]
    \centering
    \includegraphics[width=8.5cm]{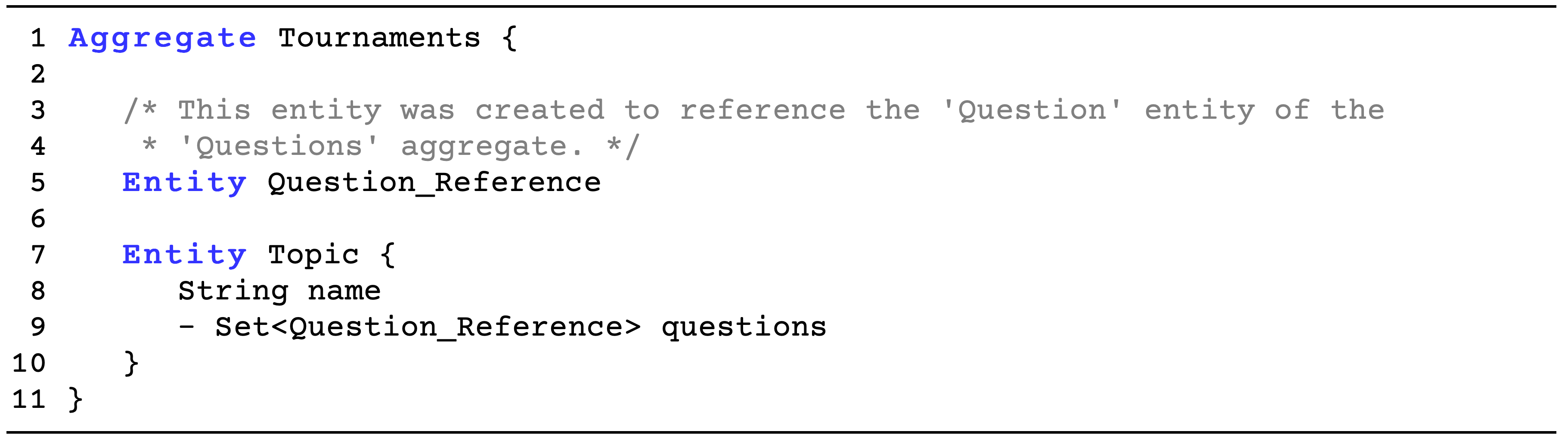}
    \caption{Generated CML example, representing an \textit{Aggregate} that contains 2 \textit{Entities}. Since \texttt{Topic} referenced an \textit{Entity} in its fields not present in the \textit{Aggregate}, \texttt{Question\_Reference} was generated locally to replace this reference.}
    \label{fig:cml_entity_reference}
\end{figure}

This problem is solved in two ways. First, when an \textit{Entity} references an outer \textit{Entity}, i.e. from an outer \textit{Bounded Context}, the \textit{Context Map} is updated with a relationship between \textit{Bounded Contexts} in the direction of the referenced \textit{Entity}. Second, this reference is replaced with a reference to a newly created local \textit{Entity}, which represents the outer \textit{Entity} locally. This is done so that the architect can better visualize which references in \textit{Entities} need to be refactored, facilitating the modeling work. Figure~\ref{fig:cml_entity_reference} shows an example of an outer reference being resolved. A comment is also automatically generated to make these references stand out.

\vspace{-5mm}

\subsubsection{Functionality Mapping} Functionalities are more challenging to represent in DDD since each functionality is composed of a sequence of read and write accesses to entities, which is a concept very particular to Mono2Micro and without apparent DDD equivalent concept. Additionally, the sequence of accesses that represents a functionality can be quite extensive. The reason for this is the fine-grained nature of the accesses collected from monolith code, due to their object-oriented design. This contrasts with the coarse-grained communication that is expected between microservices to avoid distribution communication costs. Without resolving this granularity issue, it becomes very impractical to represent functionalities compactly.

Fortunately, Mono2Micro provides a Functionality Refactoring tool that rewrites the functionalities of a candidate decomposition as Sagas~\cite{Almeida20,Correia22}. The tool converts several fine-grained microservice invocations into some coarse-grained ones, incorporating the Remote Façade pattern in the decomposition. Refactoring the functionalities using this tool has an extreme positive effect on the granularity of the access sequences, so the Functionality Refactoring module is represented in Figure~\ref{fig:m2m_pipeline_extension} as a crucial step in the extended pipeline.

Refactoring functionalities as Sagas also makes a possible map to DDD more adequate. Although the Saga pattern is not a DDD pattern, in practice it can be used in conjunction with DDD to model distributed transactions~\cite{Neal21}. Similarly, Context Mapper also supports constructs in addition to the original DDD patterns that enrich the modeling capabilities of the tool. For the case of functionality representation, the definitions of \textit{Use Case} and \textit{Event Flows} are the two most obvious constructs to use. However, they have some drawbacks. \textit{Use Case} definitions would imply reverse engineering of the functionalities as they are to a previous stage of the development cycle, losing in the process the information of the entity accesses and their Saga format. \textit{Event Flows} seem more plausible to model Sagas, but they would lock the architect into using Event-Driven Design for all functionalities, which should be left for the architect to decide later case by case instead of enforcing it through the automatic translation to CML.

To supply a construct for the representation of Sagas meeting the current requirements, an expansion to the CML syntax was proposed and implemented in Context Mapper, which allows for the definition of distributed workflows without specifying the communication model of the process. For the current functionality mapping use case, this new concept can be used to simply state the steps of the saga, without any implementing technology commitments.

This construct is called \textit{Coordination}, and is based on the coordination property of Sagas that specifies whether the steps of a Saga are orchestrated or choreographed~\cite{Neal21}, and nothing more.



\begin{figure}[htb]
    \centering
    \includegraphics[width=8.5cm]{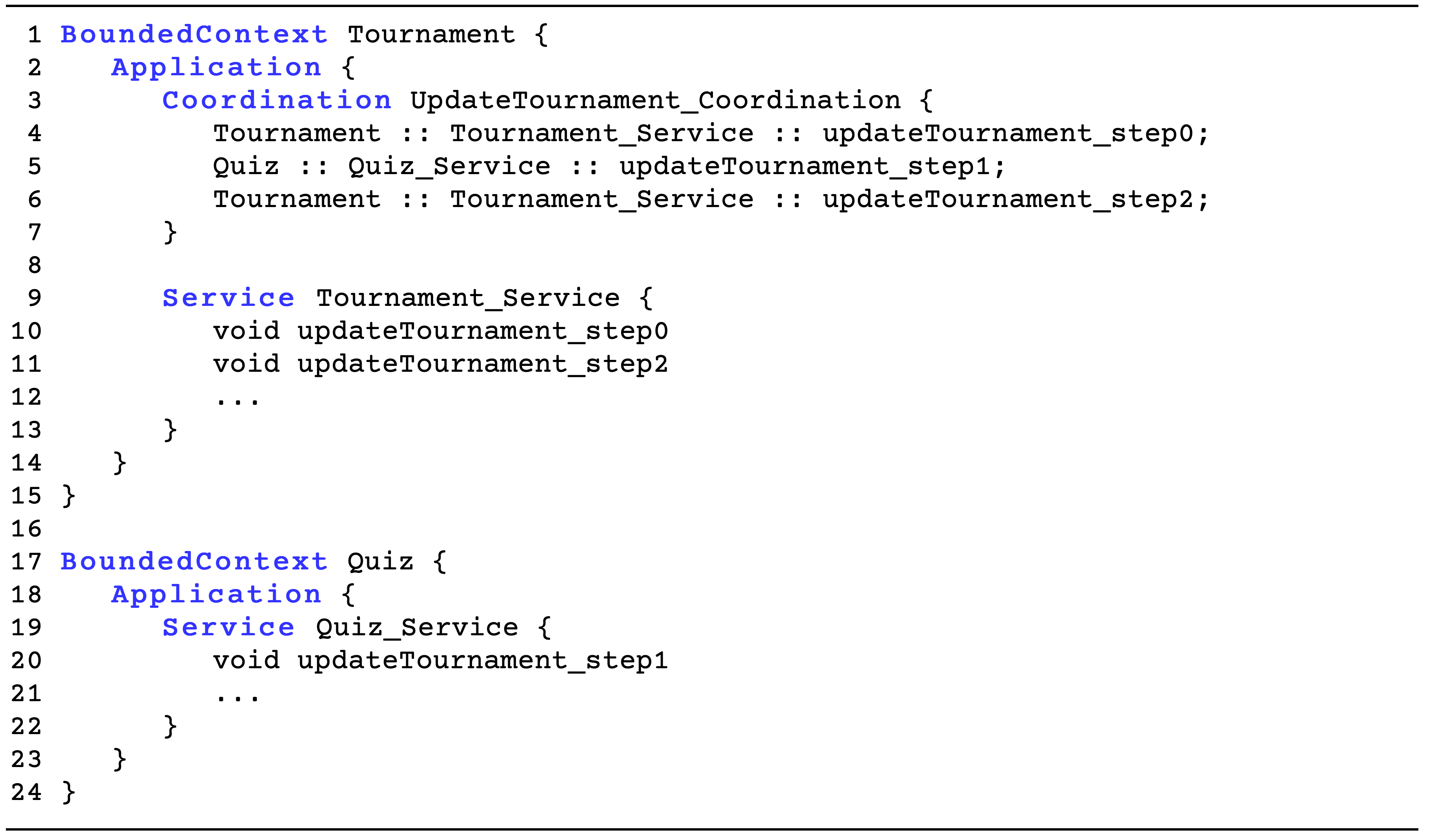}
    \caption{\textit{Coordination} construct in CML. The steps of the \textit{Coordination} (4-6) represent ordered calls to \textit{Service} operations (10,20,11).}
    \label{fig:cml_coordination}
\end{figure}

In CML, \textit{Coordinations} can be used to coordinate defined \textit{Service} operations, the same way a Saga coordinates steps. Figure~\ref{fig:cml_coordination} shows an example of the syntax in CML. \textit{Coordinations} are defined within the \textit{Application} layer of a \textit{Bounded Context}. To reference a \textit{Service} operation, a coordination step is divided into three segments, separated by the \texttt{::} symbol: The name of the Bounded Context where the operation is defined; the name of the application \textit{Service} where the operation is defined; and the name of the operation.

Functionalities that do not access other \textit{Bounded Contexts} are simply mapped to a \textit{Service}, also defined in the \textit{Application} layer of the \textit{Bounded Context} where they are defined.

\subsection{CML Representation and Interaction}
\label{subsec:cml_interaction}

When using Mono2Micro, architects now have the option to convert candidate decompositions to CML using the translation strategy mentioned so far.

Like all CML discovery strategies, the initial representation of the candidate decomposition in CML is not final. Further refactoring is expected. However, an effort was made to automatically create a good starting point. The names of entities, clusters, and functionalities from the initial decomposition are maintained and used for naming \textit{Entities}, \textit{Aggregates}, \textit{Bounded Contexts}, and \textit{Coordinations} in CML. The conversion of the format of functionalities to Sagas also creates additional constructs, in the form of \textit{Service} operation calls, which correspond to \textit{Coordination} steps in CML. These operations make up the interface of each \textit{Bounded Context} in CML, but there is no straightforward name that can be used to name each operation. As such, several access-based naming heuristics were implemented in the translation strategy. Names are formatted with the names of entities, which are prefixed by the type of access and separated by a dash. For example, assuming that the sequence of accesses of the \textit{Coordination} in Figure~\ref{fig:cml_coordination} is: \texttt{Tournament} read for step 0; \texttt{Quiz} read/write and \texttt{Question} read/write for step 1; and \texttt{Tournament} write for step 2. Each service call name would be generated as \texttt{rTournament}, \texttt{rwQuiz\_rwQuestion} and \texttt{wTournament}, respectively. The architect can further customize the level of detail they want the name to have regarding access information:

\begin{itemize}
    \item Full Access Trace: Transcribe the entire ordered entity access sequence that happens within an operation into the name of that operation;
    
    \item Ignore Access Types: Omit the type of access to entities in operation names, i.e. read/write, replacing it with a "ac" prefix;

    \item Ignore Access Order: Omit the type and order of access to entities in the operation names.
\end{itemize}

Each heuristic used reduces the number of generated operations, at the cost of entity access details. In its most reduced form, each operation name shows which entities are accessed in that step. Access information is also added to each translated entity in the form of a comment, showing metrics related to the percentage of external and local accesses to the entity in comparison with the total external and local accesses to the \textit{Bounded Context}. In contrast to these heuristics, there is also the option to generate generic names for operations that are not access-based. These names are composed of the name of the functionality and the step number in the corresponding \textit{Coordination}, meaning each step of each \textit{Coordination} will call a unique operation.

In terms of interaction, by opening this CML artifact in Context Mapper, architects have at their disposal a DDD-based view of the candidate decomposition, and several features to further refine, refactor, and visualize it. Table~\ref{tab:operation_comparison} shows some of these new features compared to what was previously available in Mono2Micro. Architects can refactor decompositions at a more structural level in CML. They can also use the additional DDD patterns present in CML to add to the initial translation, especially strategic ones, to define, for example, what type of relationships exist between the translated \textit{Bounded Contexts}.

\begin{table}
    \footnotesize
    \centering
    \caption{Operations in Mono2Micro and Context Mapper.}
    \begin{tabular}{c c}
        \toprule
        \textbf{Mono2Micro} & \textbf{Context Mapper}\\
        \midrule

        Expand / Collapse cluster views & Merge / Split Bounded Contexts \\

        Merge / Split clusters & Merge / Split Aggregates \\

        Transfer entities between clusters & Design Bounded Context Relationships \\

        Rename clusters & Redesign Coordinations \\

        Redesign functionalities & Redesign Entities \\

        Analyse access-based measures & Generate UML / BPMN representations \\
        \bottomrule
    \end{tabular}
    \label{tab:operation_comparison}
\end{table}

\section{Case Study}
\label{sec:case_study}

Quizzes-Tutor (QT)\footnote{\url{https://quizzes-tutor.tecnico.ulisboa.pt/}} is an online quizzes management application developed for educational institutions. It can be used to create, manage, and evaluate quizzes composed of varying types of question formats. Teachers can add questions related to topics of the courses they preside over, while students can answer these questions within quizzes. Other functionalities include the creation of quiz tournaments between students, question proposals from students, and ways to discuss question answers. This real-world monolith, composed of 46 domain entities and 107 functionalities, was used as a case study to validate the Mono2Micro pipeline extension, which provides DDD modeling capabilities.

\subsection{Decomposition Generation}
\label{subsec:decomposition_generation}

\begin{figure}[htb]
    \centering
    \includegraphics[width=6.5cm]{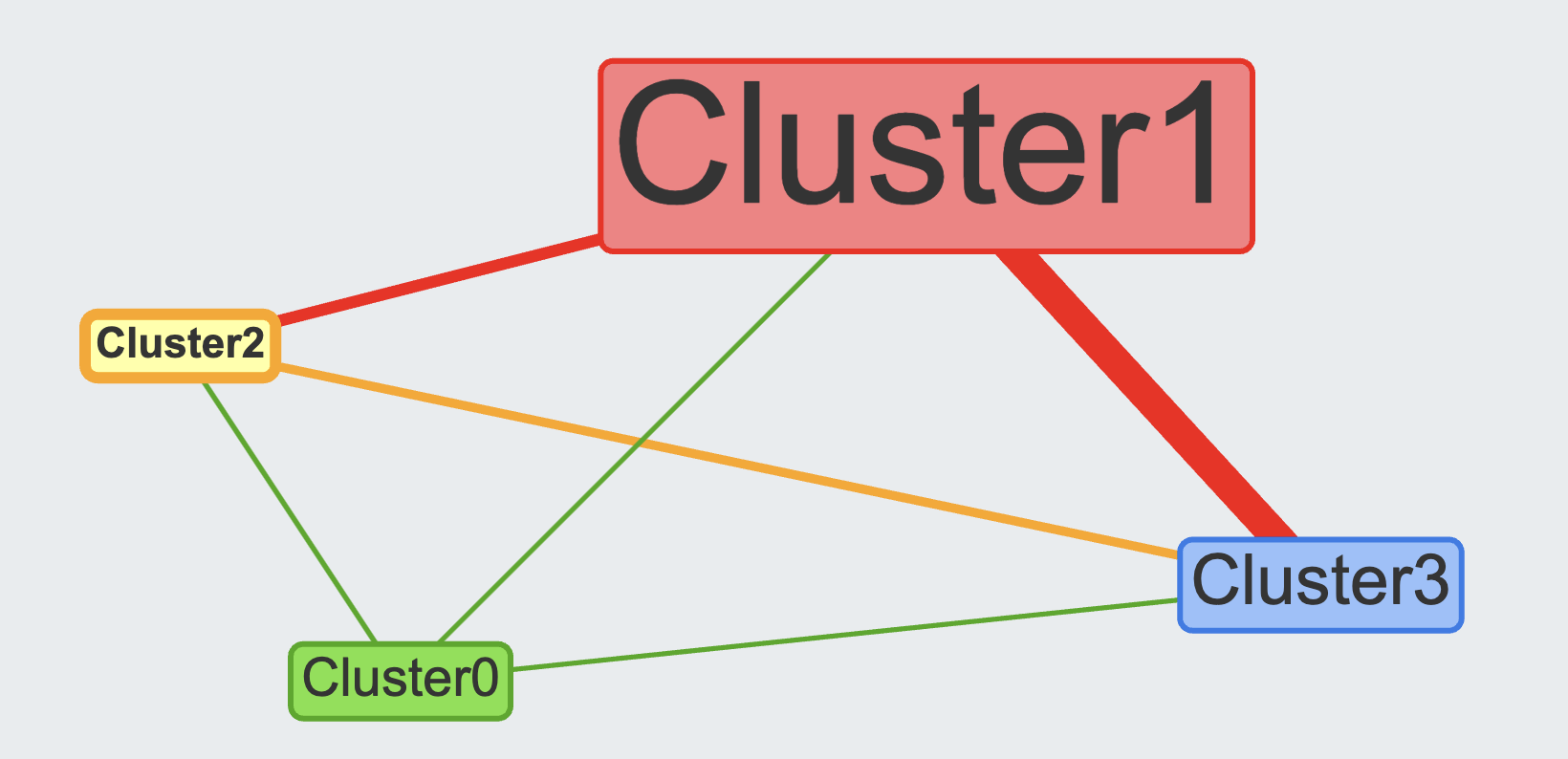}
    \caption{Mono2Micro decomposition visualization with fine-grained interaction between clusters. Edges represent functionalities shared between clusters.}
    \label{fig:m2m_decomposition_view}
\end{figure}

To start the validation, a candidate decomposition for the QT application must be generated and chosen. To this end, around 2000 candidate decompositions were generated with different values of similarity criteria and number of clusters. Candidate decompositions were then filtered on the basis of the values of their measures. The heuristic used was to order decompositions based on their coupling value in ascending order, and then based on their cohesion value in descending order, to prioritize decompositions with low coupling and high cohesion. Of the top 100 results, the candidate decomposition with the lowest complexity value was chosen. Data for this candidate decomposition can be seen in Table~\ref{tab:decomposition_results}. Figure~\ref{fig:m2m_decomposition_view} also shows the clusters view of the decomposition shown in Mono2Micro.

\begin{table}
    \scriptsize
    \centering
    \caption{Candidate decomposition measures for the QT case study.}
    \begin{tabular}{c c c c c c}
        \toprule
        \textbf{Cluster} & \textbf{Entities} & \textbf{Functionalities} & \textbf{Cohesion}& \textbf{Coupling}& \textbf{Complexity}\\
        \midrule

        Cluster0 & 6 & 7 & 0.81 & 0.185 & 787.571 \\
         
        Cluster1 & 27 & 107 & 0.212 & 0.657 & 106.832 \\
         
        Cluster2 & 4 & 11 & 0.727 & 0.179 & 431.091 \\
         
        Cluster3 & 9 & 35 & 0.654 & 0.753 & 322.486 \\
        \bottomrule
    \end{tabular}
    \label{tab:decomposition_results}
\end{table}

Table~\ref{tab:decomposition_results} shows two noteworthy pieces of information. To start with, the complexity of each cluster is very high. The complexity measure represents the migration cost of the functionalities in a cluster. This migration cost is measured as the cost of re-designing from an ACID context to a distributed one. Of the initial 107 functionalities, 31 involve distributed calls composed of several hops between clusters that drive the complexity high. This is because functionalities are still represented by the fine-grained monolith interactions between entities that can now be in different clusters. To reduce this complexity, the \textit{Functionality Refactoring} tool represented in Figure~\ref{fig:m2m_pipeline_extension} is used to create coarse-grained interactions between clusters. Table~\ref{tab:complexity_reduction} shows the reduction of invocations for some of the QT functionalities. Applying this complexity reduction also makes it viable for functionalities to be represented in a structural language such as CML. Otherwise, any translation strategy would culminate in thousands of operation definitions for just a subset of the functionalities, as the FGI (Fine-Grained Interaction) values show in Table~\ref{tab:complexity_reduction}.

\begin{table}
    \footnotesize
    \centering
    \caption{Refactored functionalities for QT case study. CGI stands for Coarse-Grained Interaction, and FGI stands for Fine-Grained Interactions.}
    \begin{tabular}{c c c c c}
        \toprule
        \textbf{Name} & \textbf{\#Clusters} & \textbf{CGI} & \textbf{FGI} & \textbf{Reduction\%}\\
        \midrule

        concludeQuiz & 3 & 4 & 73 & 94.52\\
         
        getQuizByCode & 3 & 4 & 33 & 87.88\\
         
        getQuizAnswers & 4 & 8 & 84 & 90.48\\
         
        exportCourseExecutionInfo & 4 & 9 & 110 & 91.82\\
         
        importAll & 3 & 5 & 119 & 95.8\\
         
        createQuestion & 2 & 3 & 24 & 87.5\\
         
        getQuizAnswers & 4 & 8 & 92 & 91.30\\
        \bottomrule
    \end{tabular}
    \label{tab:complexity_reduction}
\end{table}

The other noteworthy piece of information is the high number of entities inside Cluster1 compared to the other clusters, which means that the entities inside this cluster are more entangled when it comes to the functionalities that use them, and are more difficult to separate without creating an overly complex decomposition. It is also the reason for the non-optimal levels of cohesion and coupling in this cluster. At this stage, when some manual refactoring of functionalities is needed, modeling using DDD can help.

\begin{figure}[htb]
    \centering
    \includegraphics[width=8.5cm]{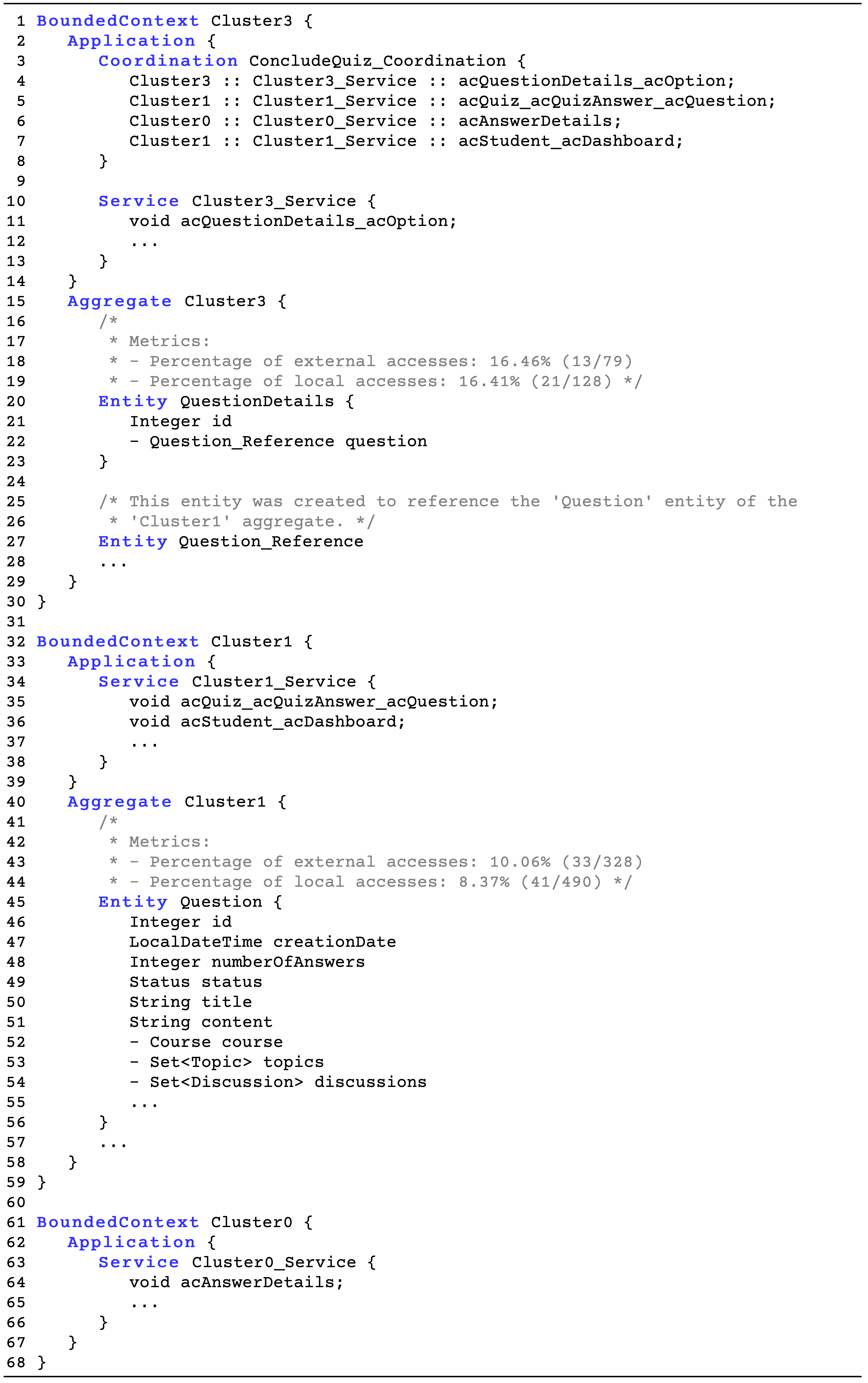}
    \caption{Snippet of the generated CML related to the functionality \texttt{ConcludeQuiz}. Triple dots (...) represent omitted constructs for the purpose of the example. Service operation names were also truncated.}
    \label{fig:case_study_cml}
\end{figure}

The candidate decomposition is translated into CML by the discovery strategies, which outputs a \texttt{.cml} file with a representation of the candidate decomposition. Figure~\ref{fig:case_study_cml} shows a modified snippet of the generated CML, related to the \texttt{ConcludeQuiz} functionality of a decomposition. Without any optimization, the translation generates a total of 121 operations, used by 31 Coordinations that represent the distributed functionalities. With naming heuristics, the number of service calls can be reduced to 87 using \textit{Full Access Trace}, to 84 using the \textit{Ignore Access Types}, and to 48 using the \textit{Ignore Access Order}. Table~\ref{tab:service_reduction} shows the reduction of generated services according to which heuristics are used per cluster.

Regarding entity generation, a total of 11 reference entities were generated to signal structural dependencies between entities, also shown in Table~\ref{tab:service_reduction}, and every entity is generated with information on the number of accesses to it, from the total \textit{Bounded Context} accesses (external and local), as shown in Figure~\ref{fig:case_study_cml}. This information can help to refactor the decomposition further. For example, in Cluster3, there are 2 dominant entities out of 9 in terms of external accesses: \texttt{QuestionDetails} and \textit{Image}. These make them good possible candidates to serve as \textit{Aggregate Roots} to the \textit{Bounded Context}. A look at their structural relationships reveals that \texttt{QuestionDetails} also contains 3 subclass entities in the same cluster: \texttt{MultipleChoiceQuestion}, \texttt{CodeOrderQuestion} and \\ \texttt{CodeFillInQuestion}; which in turn are structurally related to: \texttt{Option}, \texttt{CodeOrderSlot}, \texttt{CodeFillInSlot} and \texttt{CodeFillInOption}. \texttt{Image} is not structurally related to these, but its presence in this \textit{Bounded Context} means its commonly accessed together with the other entities. A possible first refactoring of this cluster could then be the separation of the Cluster3 \textit{Aggregate} into 2 \textit{Aggregates}, which is an automatic architectural refactor supported by CML. The first \textit{Aggregate} would control question types, formats, and related invariants with \texttt{QuestionDetails} as its \textit{Aggregate Root}. The second \textit{Aggregate} would represent a repository of images and be controlled through its \texttt{Image} \textit{Aggregate Root}.

\begin{table}
    \small
    \centering
    \caption{Generated CML constructs. The number of services is represented by four values: No heuristics used; \textit{Full Access Trace} used; \textit{Ignore Access Types} used; and \textit{Ignore Access Order} used. The number of entities is represented by two values: original entities and reference entities. The most accessed entity is based on external accesses to the \textit{Bounded Context}.}
    \begin{tabular}{c c c c}
        \toprule
        \textbf{Cluster} & \textbf{\#Services} & \textbf{\#Entities} & \textbf{Most Accessed Entity}\\
        \midrule

        Cluster0 & 15/7/6/4 & 6/4 & QuizAnswerItem (35.14\%)\\

        Cluster1 & 59/53/52/34 & 27/3 & Quiz (12.2\%)\\

        Cluster2 & 11/5/5/2 & 4/0 & QuestionAnswerItem (30.0\%)\\

        Cluster3 & 36/22/21/8 & 9/4 & QuestionDetails/Image (16.46\%)\\
    \end{tabular}
    \label{tab:service_reduction}
\end{table}

\section{Discussion}
\label{sec:discussion}

This section discusses the findings of applying the DDD-based extension to the operational pipeline of Mono2Micro by analyzing how the implemented solution and the results of its application in the case study answer the research questions raised in the introduction of this paper.

\subsection{Results Validation}
\label{sec:results_validation}


Starting with the first research question (RQ1), to understand whether the Mono2Micro operational pipeline can be extended to integrate DDD, through the use of CML, the first step taken was to compare the architectural characteristics of each tool. 

Mono2Micro follows a pipe-and-filter architecture with its stages, and provides extension points in the form of abstractions for each stage; while Context Mapper follows a hub-and-spoke architecture, and also provides extension points in its core, the CML language, and each of its spoke modules, including the Discovery Library (DL). As described in the solution, these characteristics show that the tools emphasize modularity and extensibility, which made the development of an integration strategy that follows the same requirements possible. By measuring the level of modularity and extensibility of the solution, RQ1 can be evaluated.

First, modularity deals with how divided a system is into logical modules that encapsulate specific self-contained functionality, improving separation of concerns and internal cohesion. The solution is composed of two new modules in Mono2Micro, the \textit{Structure Collector} and \textit{CML Translator}. In terms of cohesion, both modules respect the pipeline architecture of Mono2Micro, and are placed accordingly inside it based on their responsibilities. Regarding coupling, Figure~\ref{fig:m2m_pipeline_extension} also shows the dependencies of each model. The \textit{Structure Collector} is highly decoupled, depending only on the monolith source code. It does not know about the subsequent stages of the pipeline. The \textit{CML Translator} has a contract between it and the Mono2Micro model, ensuring that it is only coupled to one source.

Second, extensibility deals with how open for extension the features of a system are without putting at risk their core structure, improving the addition of new functionality. The \textit{Structure Collector} was designed from scratch. It provides abstractions for the collection of data from new frameworks and the collection of other types of structural data. The \texttt{CML Translator} is an extension of the DL API, so it follows that the discovery strategies implemented follow the same design and are also open to extension by providing abstractions. Additionally, the contract generation is also built with abstractions if new information is needed for future discovery strategies.


Moving on to the second research question (RQ2), the mappings of the cluster, entity, and functionality concepts to DDD demonstrate how a candidate decomposition based on entity accesses can be represented with DDD concepts. Mono2Micro entities are already based on the concept of DDD \textit{Entities}, so the mapping is consistent in this regard. Further more, to improve usability, structural information about entities is collected for a more complete representation in CML. In the case of clusters, consistency was maintained by mapping each of them to a \textit{Bounded Context} and an \textit{Aggregate}. The structural context of the CML was also considered. To avoid having the cluster entities break the bounds of outer \textit{Bounded Contexts} with direct references, reference entities were added. This improves usability by highlighting where refactoring needs to be applied to entity relationships. For the mapping of functionalities, the sequence of accesses to entities that composed them was first converted into a structured Saga. This significantly reduced the complexity of the sequence in terms of size and hops between clusters, and made it simpler to represent with DDD, since Saga is a well-known pattern in distributed systems. Sagas were mapped to \textit{Coordinations} in CML, which encode an ordered sequence of service calls, just as Sagas encode a sequence of steps. To improve the usability of this mapping, hints to which entities are accessed in each service call were generated in the names of each service call.


Finally, in regard to the third research question (RQ3), analyzing and comparing the decomposition artifacts represented in each tool, as demonstrated in the case study, can show how an architect can benefit from the use of this extension.

In the case of entity representation, structural information is now available in the CML representation. This information is novel in regard to the old pipeline, as Mono2Micro representations can at most show the names of entities within each cluster, not its internal state or structural connections, as illustrated in the QT decomposition in Figure~\ref{fig:m2m_decomposition_view}. With the CML representation of entities, it is possible to observe the attributes of each entity and also the structural refactorings that must be made in existing entities based on reference entities to other Bounded Contexts, demonstrated in Figure~\ref{fig:case_study_cml}.

In the case of clusters, \textit{Aggregates} can now be defined and used to further partition a cluster and its entities based on access patterns, access percentages, and the structural information provided at generation time in the form of commented constructs.

\begin{figure}[htb]
    \centering
    \includegraphics[width=5.5cm]{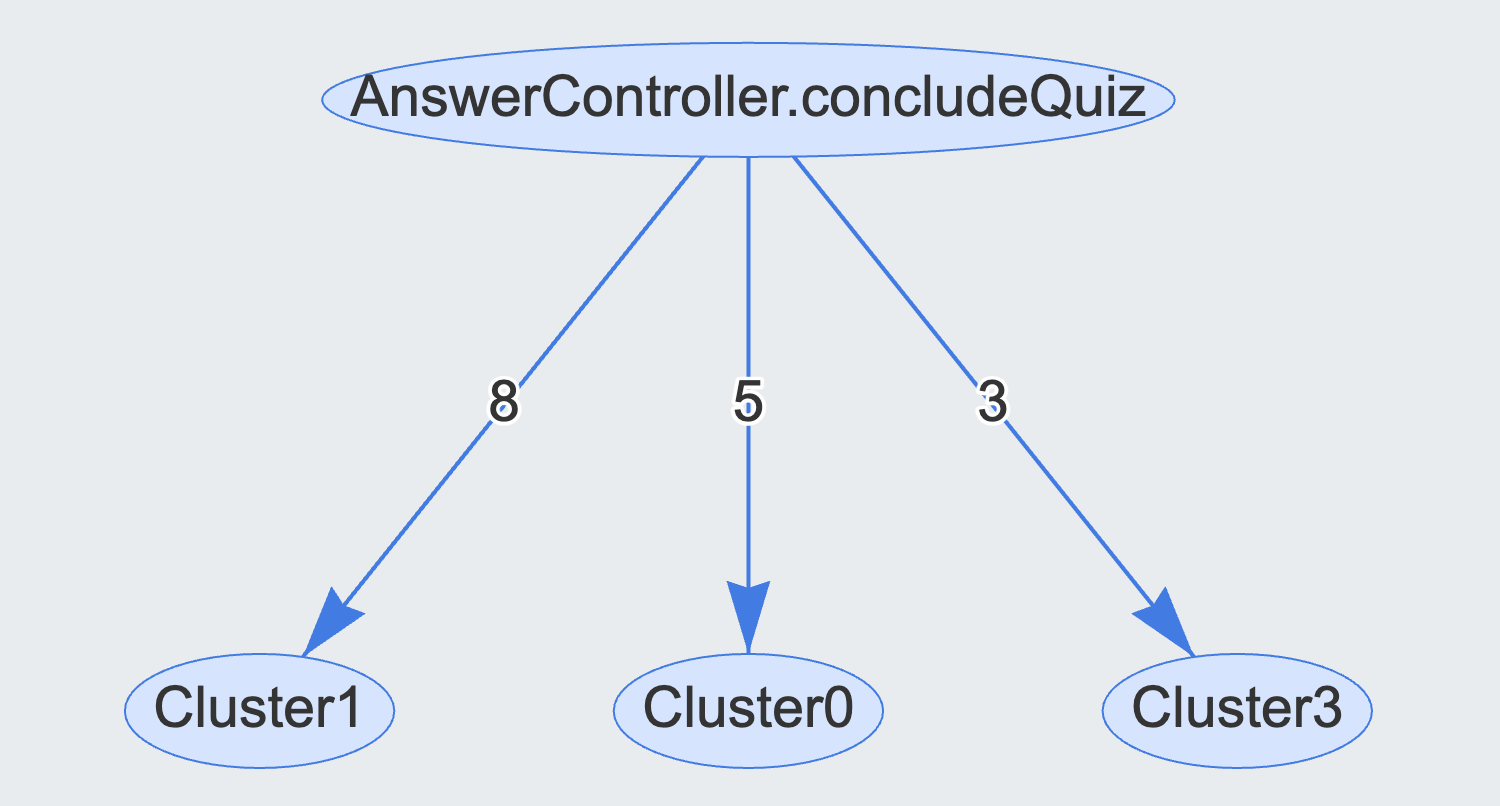}
    \caption{Graph representation of the \texttt{ConcludeQuiz} functionality in Mono2Micro, and the clusters that participate in it. Edge numbers represent the number of accessed entities in the clusters they point to.}
    \label{fig:diagram_comparison_m2m}
\end{figure}

\begin{figure}[htb]
    \centering
    \includegraphics[width=7.5cm]{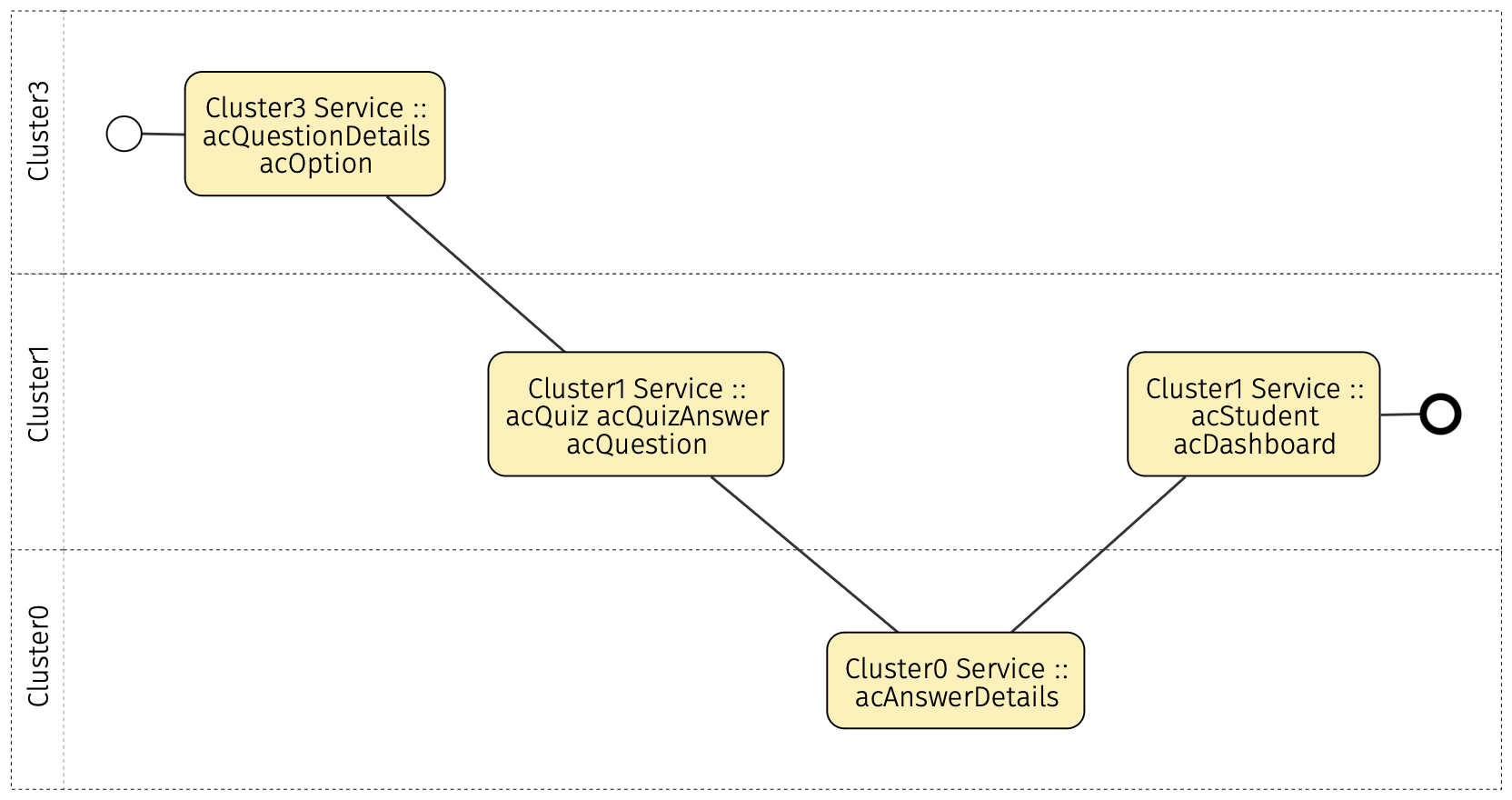}
    \caption{BPMN representation of the \texttt{ConcludeQuiz} \textit{Coordination} in Context Mapper, generated using BPMN Sketch Miner. The tasks in the diagram represent the steps of the \textit{Coordination}, and each participant (lane) represents the \textit{Bounded Context} where the step is defined.}
    \label{fig:diagram_comparison_bpmn}
\end{figure}

In the case of functionalities, the architect now has the option of editing their Saga representation in CML, by editing the generated \textit{Coordinations}. Mono2Micro only allowed for the visualization of fine-grained functionality traces in a graph representation, and the Functionality Refactoring module of the tool only produced the data for the Sagas, without any way to edit or visualize them in a graphical representation. Using CML, these functionalities can be modeled as \textit{Coordinations} and edited in the language with the added context of the internal structure of the \textit{Bounded Contexts} they interact with. Furthermore, \textit{Coordinations} can be visualized in BPMN using a prompt to generate the diagram in BPMN Sketch Miner\footnote{https://www.bpmn-sketch-miner.ai}. Figures~\ref{fig:diagram_comparison_m2m} and~\ref{fig:diagram_comparison_bpmn} show the comparison of the views of a functionality that interacts with three of the four clusters before and after integration with CML.

Additionally, the service naming heuristics allow the architect to reduce the number of generated service calls that exist in each cluster. This does not reduce the number of functionality steps but increases the level of reuse of services.

\vspace{-5mm}

\vspace{-2mm}

\subsection{Practical Relevance and Adoption}
\label{sec:practical_relevance_adoption}

The problem addressed with the presented approach is of high relevance to practitioners, mainly software engineers and architects, who need to modernize existing monolith applications. Such monolith applications without an inner modular structure, a.k.a. \textit{Big Ball of Mud}~\cite{Vernon16}, turned out to be a huge challenge in practice for several reasons, which can also be seen as use cases for our solution presented in this paper:

\begin{itemize}
    \item \textbf{Economic reasons}: Maintaining a \textit{Big Ball of Mud} is often becoming expensive for software companies. Changing such applications, adding new features, or fixing bugs, often takes too much time because of the intertwined code base and dependencies within the system.
    \item \textbf{Scalability and "Cloud readiness"}: Many companies have to decompose their applications to migrate to the cloud. The monolithic architecture approach is not scalable and does not fit the requirements for cloud deployment.
    \item \textbf{Autonomous teams and "DevOps"}: Many companies aim to implement agile development approaches in which teams develop and operate their part of an application autonomously \cite{Tune17}. A team should be able to make its own design and architectural decisions, ideally without many dependencies towards other teams. This requires loosely coupled (micro-)services or at least a system with loosely coupled modules. The structure of the organization (teams) defines the architecture of the software.

\end{itemize}

As already mentioned, the adoption of DDD for service decomposition is widespread in the software industry. Both our tools, Mono2Micro\footnote{https://github.com/socialsoftware/mono2micro} as well as Context Mapper\footnote{https://github.com/ContextMapper} are open-sourced and, at least individually, have already gained some adoption in industry and real-world projects. The proposed approach is therefore foreseen as an important contribution and support for practitioners who want to: use a tool that automatically suggests decompositions for an existing monolith system; and want to express their future architecture and service decomposition in terms of DDD patterns and follow the "domain-driven" approach. Once a CML model is available, practitioners can benefit from all Context Mapper features: iterative and agile modeling, architectural refactorings, model visualization (diagram generators), or even code generation.

\subsection{Threats to Validity}
\label{sec:threats_to_validity}


With respect to internal validity, there are two points worth considering. First, the functionalities used in the decomposition and CML mapping process are all linear in nature. This is due to the existent static entity access collection tool in Mono2Micro, which flattens code branches into a single access sequence in a depth-first fashion. However, previous research that used the same sequences to develop the Saga representation of functionalities has shown this has little impact on the final results~\cite{Correia22}, and support for multiple traces per functionality is being developed. By extension, the current implementation of \textit{Coordinations} in CML is also based on linear Sagas. However, \textit{Coordinations} can be opened in BPMN Sketch Miner, where they can be further edited with more complex workflow logic. Context Mapper also supports branching processes in its \textit{Event Flows}, so the concept could be expanded to \textit{Coordinations} as well in the future. 

Second, the provided solution for resolving service operation names with access traces can sometimes generate verbose names, but these are meant to be temporary and only inform on what a certain step is doing, which is more appropriate and informative than the alternative of generating incremental "step" names. Future work could expand on the level of collected information to create heuristics based on method names to reduced verbosity.

In terms of external validity, the current implementation assumes the use of Java and the Spring Boot JPA Framework to collect entity access and structure information, but the process is general enough to be applicable to other programming languages and frameworks. The modules that assume these limitations are also built with abstractions for the implementation of other technologies.

The implementation of the solution is also realized by using Mono2Micro and Context Mapper, but this does not mean that a more general solution cannot be derived for other tools. Mono2Micro itself is designed as a generalization of the state of the art in microservice identification~\cite{Lopes23}. As for Context Mapper, the tool is built on top of the concepts and patterns of tactical and strategic DDD. If other tools follow representations of the same patterns, they can be used by the same mapping strategies applied to Context Mapper.

%


\section{Conclusion}
\label{sec:conclusion}

A significant amount of research has been done on the migration of monoliths into a microservice architecture, but almost no tools incorporate mappings to DDD concepts in their migration processes. Research showed a trend for developing DSLs to represent DDD concepts and adapting the concepts to work in other diagramming tools to develop microservices, but to our knowledge never directly in a migration tool.

Practitioners and software architects who need to migrate their current monolith architectures due to economic reasons, scalability, or more autonomous development teams would benefit from a solution that not only proposes candidate decompositions, but also automatically generated design-level DDD artifacts from that decomposition, as a starting point to facilitate the refactoring process towards a microservice architecture.

This paper proposes a solution pipeline for this problem composed of the integration of Context Mapper, a modeling framework that provides a DSL to represent DDD patterns, into the Mono2Micro decomposition pipeline, a robust microservice identification tool.

The proposed solution achieves the integration of both tools by defining a mapping of concepts between tools, whilst respecting each of the tool models. To support this mapping, the solution includes several new modules and modifications to the tools, including a new static collector of entity structural information, a contract for effective communication between the tools, a translation strategy to generate CML from Mono2Micro decompositions, i.e. entities, clusters, and functionalities, an extension to the CML syntax to support concepts from decomposition in the form of \textit{Coordinations}, and new diagram generators from CML based on translated decompositions.

The artifacts developed in the project are publicly\footnote{\url{https://github.com/socialsoftware/mono2micro/tree/master/tools/cml-converter}} available together with the description of the procedures necessary to use them.



\begin{acks}

This work was partially supported by Fundação para a Ciência e Tecnologia (FCT) through projects UIDB/50021/2020 (INESC-ID) and PTDC/CCI-COM/2156/2021 (DACOMICO)
\end{acks}

\bibliographystyle{ACM-Reference-Format}
\bibliography{sample-base}


\end{document}